\documentstyle[twoside,epsf]{article}

\input{ibvs2.sty}

\begin{document}

\IBVShead{xxxx}{xx xxx 2003}

\IBVStitle{VY S\lowercase{cl}-Type Star V504 C\lowercase{en}}

\IBVSauth{Kato,~Taichi$^1$; Stubbings,~Rod$^2$}
\vskip 5mm

\IBVSinst{Dept. of Astronomy, Kyoto University, Kyoto 606-8502, Japan,
          e-mail: tkato@kusastro.kyoto-u.ac.jp}

\IBVSinst{19 Greenland Drive, Drouin 3818, Victoria, Australia,
       e-mail: stubbo@sympac.com.au}

\IBVSobj{V504 Cen}
\IBVStyp{NL}
\IBVSkey{cataclysmic variable, VY Scl-type star, photometry}

\begintext

   V504 Cen is classified as a possible R CrB-type star in General Catalogue
of Variable Stars (Kholopov et al. 1985).  Almost nothing has been
published on its nature of variability.  Kilkenny and Lloyd Evans
(1989), during the course of a systematic study of R CrB-type candidates,
noticed that V504 Cen shows broad Balmer emission lines.  From this
spectroscopic feature, Kilkenny and Lloyd Evans (1989) concluded that
V504 Cen is a cataclysmic variable (CV), and not a R CrB-type star.
Although Kilkenny and Lloyd Evans (1989) proposed that this star may
be a VY Scl-type CV, which is characterized by the presence of occasional
deep fadings, the lack of long-term light curve with sufficient limiting
magnitudes made the classification rather inconclusive.  In spite of
this report, the object has been largely neglected from the past CV
research.  The object has not been listed in any comprehensive CV
catalogs, including the latest edition of the CV catalog by
Downes et al. (2001).

   We here present first-ever published light curve of V504 Cen,
which clearly demonstrates the VY Scl-type nature.  The light curve is
drawn from visual observations (RS) and ASAS-3 $V$-band public data
(Pojmanski 2002).  A 0.45 mag systematic correction was added to RS's
visual observations based on the RASNZ comparison stars, in order to
best match the ASAS-3 $V$ magnitudes.  Figure 1 shows the entire
light curve between 1998 and 2003.  A deep fading episode between
JD 2452346 and 2452650 is apparent.  The fading portion of this
fading episode was not recorded because of the solar conjunction.
Although not shown in the light curve, 49 ASAS-3 observations between
JD 2452405 and 2452705 gave only negative detections.  The object
must have been fainter than $V$=13.6 during this period.

\IBVSfig{12cm}{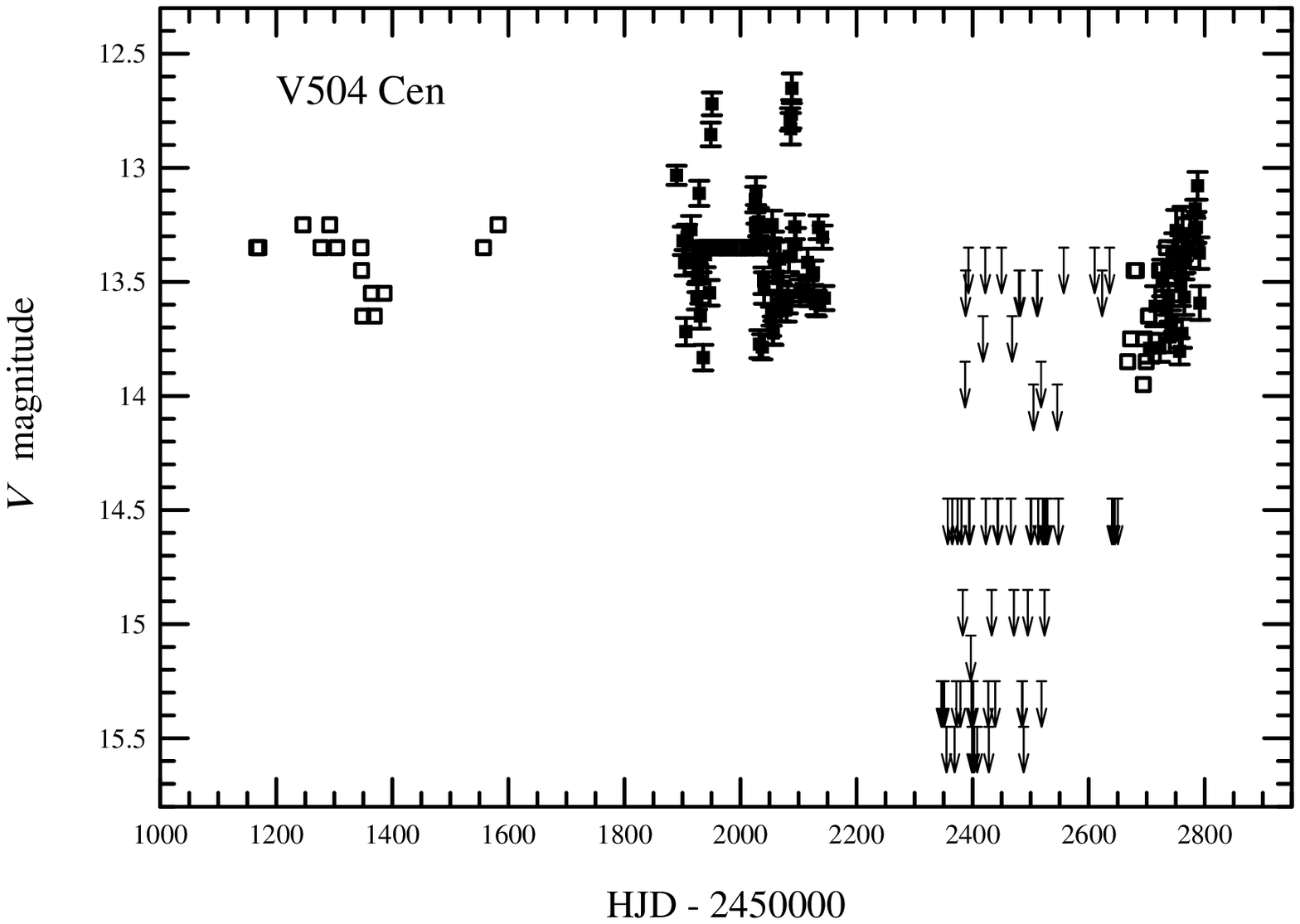}{Light curve of V504 Cen from RS's visual
observations and the ASAS-3 $V$-band public data.
The filled squares with error bars and open squares represent
ASAS-3 data and RS's measurements, respectively.  (The straight
bar between JD 2451936 and 2452041 represents RS's observations).
The arrows represent upper limit observations by RS.
}

\IBVSfig{10cm}{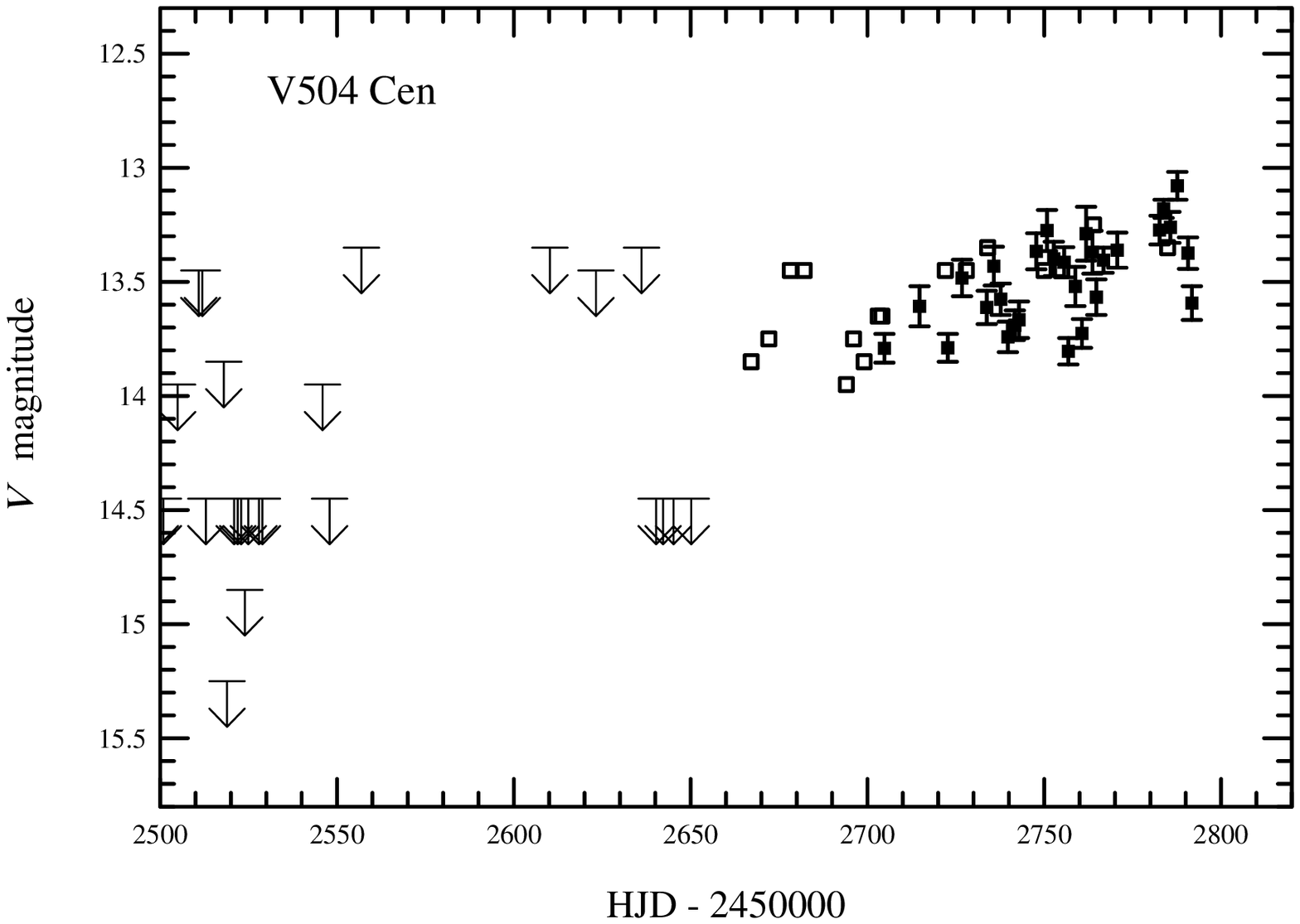}{Enlarged light curve of the recovery stage
from the deep fading.  The symbols are the same as in Figure 1.
}

   The duration ($>$300 d, $<$400 d) and depth ($>$2.0 mag) are
typical for VY Scl-type fadings (Greiner 1998).  Figure 2 shows the
enlarged light curve of the rising portion observed in early 2003.
After reaching $V$=13.7, the object slowly returned to its maximum
magnitude.  The rising rate between JD 2452714 and 2452784 is
0.005 mag d$^{-1}$.
Such a slow final rise to the maximum is characteristic
to the VY Scl-type phenomenon (Greiner 1998; Kato et al. 2002).
In all aspects, V504 Cen is now firmly classified as a typical
VY Scl-type CV.

   Since there is no finding chart of V504 Cen readily available,
we note the following identification: V504 Cen = GSC 7808.1570
located at 14\hr 12\mm 49\fsec 11, $-$40\deg 21\arcm 37\farcs 1
(J2000.0, GSC 1.2).

\vskip 5mm

We are grateful to G. Pojmanski for making the ASAS-3 survey data
publicly available, and generously allowing us for unlimited usage.
This work is partly supported by a grant-in-aid (13640239, 15037205)
from the Japanese Ministry of Education, Culture, Sports, Science and
Technology.

\references

Downes, R. A., Webbink, R. F., Shara, M. M., Ritter, H., Kolb, U.,
  Duerbeck, H. W., 2001, PASP, 113, 764

Greiner, J., 1998, A\&A, 336, 626

Kato, T., Ishioka, R., Uemura, M., PASJ, 54, 1033

Kholopov, P. N., Samus', N. N., Frolov, M. S., Goranskij, V. P.,
  Gorynya, N. A., Kireeva, N. N., Kukarkina, N. P.,
  Kurochkin, N. E. et al., 1985, General Catalogue of Variable Stars,
  fourth edition (Moscow: Nauka Publishing House)

Kilkenny, D., Lloyd Evans, T., 1989, Observatory, 109, 85

Pojmanski, G. 2002, Acta Astronomica, 52, 397

\endreferences

\end{document}